\documentclass[onecolumn,showpacs,floatfix,11pt,nofootinbib]{revtex4}
\usepackage[dvips]{graphics,color}
\usepackage{epsfig}\usepackage{float}
\RequirePackage{amssymb}
\usepackage{makeidx}
\usepackage[brazil]{babel}
\usepackage[latin1]{inputenc}
\usepackage{graphicx}
\usepackage{indentfirst}
\usepackage{color}
\newcommand{\be}{\begin{equation}}
\newcommand{\ee}{\end{equation}}

\newcommand{\ba}{\begin{eqnarray}}
\newcommand{\ea}{\end{eqnarray}}

\begin{document}

\title{Positive tension 3-branes in an $AdS_{5}$ bulk}

\author{M. C. B. Abdalla$^{1}$}
\email{mabdalla@ift.unesp.br}
\author{M. E. X. Guimarães$^{2}$}
\email{emilia@if.uff.br}
\author{J. M. Hoff da Silva$^{3}$}
\email{hoff@feg.unesp.br}

\affiliation{1. Instituto de F\'{\i}sica Te\'orica, Universidade
Estadual Paulista, R. Dr. Bento Teobaldo Ferraz 271 - Bl. II -
Barra Funda 01140-070 S\~ao Paulo, SP, Brazil}

\affiliation{2. Instituto de F\'{\i}sica, Universidade Federal
Fluminense, Av. Gal. Milton Tavares de Souza, s/n - Campus da
Praia Vermelha 24210-346 Niter\'oi, RJ, Brazil}

\affiliation{3. UNESP - Campus de Guaratinguet\'a - DFQ, Av. Dr.
Ariberto Pereira da Cunha, 333 CEP 12516-410, Guaratinguet\'a-SP,
Brazil.}


\begin{abstract}
In this work, we review and extend the so-called consistency
conditions for the existence of a braneworld scenario in arbitrary
dimensions in the Brans-Dicke (BD) gravitational theory. After
that, we consider the particular case of a five-dimensional
scenario which seems to have phenomenological interesting
implications. We show that, in the BD framework, it is possible
to achieve necessary conditions pointing to the possibility of
accommodating branes with positive tensions in an AdS bulk by the presence of the
additional BD scalar field, avoiding in this way the necessity of
including unstable objects in the compactification scheme.
Furthermore, in the context of time variable brane tension, it is
shown that the brane tension may change its sign, following the
bulk cosmological constant sign.
\end{abstract}
\maketitle

\newpage

\section{Introduction}

In the basic framework establishment of the string theory, two
main outputs are noticeably delineated: the necessity of extra
dimensions and the presence of a scalar field (in the low energy
limit) sharing the action of the gravitational interaction with the usual rank-two tensorial field \cite{COR}. The advent of the braneworld scenarios, at least in its modern fashion, may also be
partially charged to the string theory developments.

In the Randall-Sundrum braneworld scenario \cite{RSI}, which can
be understood as an effective Horava-Witten \cite{HW}
compactification model, the branes are performed by two mirror
warped domain walls embedded into a five-dimensional bulk
spacetime and placed at the end of an orbifold extra dimension.
The bulk has a negative cosmological constant (AdS slice) and the
two mirror branes have tensions of opposite signs. The brane
which mimics our universe is endowed with an exponencial warp
factor (conformal to the four dimensional Minkowski metric) that
is responsible by the hierarchy problem solution.

Most of the characteristics outlined in the previous paragraph may
be reproduced by a five-dimensional particularization of a quite
ingenious reformulation of the Einstein tensor components,
redesigned in order to take the existence of branes
and extra dimensions into account \cite{CON,MAIS}. Here we shall apply the
constraints emerging from that program (the search for the consistency
conditions) to the case of three branes embedded into a
five-dimensional bulk spacetime within the Brans-Dicke theory.
Apart of the mentioned motivation coming from advances in String
Theory and other unification programs (such as supergravity
and M-theory), there are strong evidences that General Relativity
is a natural attractor of more general scalar-tensor gravity
theories \cite{DAM}, from which the Brans-Dicke case is just the
simplest viable example. An important result obtained from such an
extension of the consistency conditions is that the Brans-Dicke
scalar field relax the obtained constraints, allowing the possibility
of an AdS bulk only if the brane tensions are positive. I. e., it seems to be unnecessary to include a
negative brane tension {\it even in the five-dimensional bulk
case}. This is a quite remarkable result, since negative tension
branes are unstable objects. Certainly each particular model must be investigated case by case,
but the results emerging from the consistency conditions point to scenarios
without the need for negative tension branes.

The possibility of a compactification scheme with only positive
tension branes is allowed in the General Relativity theory in two
cases: 1) more than five dimensions (being six dimensions
enough for this purpose) and 2) with the presence of bulk tachyon
matter. In the first case, the consistency conditions are relaxed,
and the necessity of including a negative tension
brane is not required. In the last case, the presence of a non-canonical scalar
field in the bulk modifies the junction conditions, allowing an
additional freedom in the brane tensions sign \cite{INDY}.

Roughly speaking, although even in the five-dimensional bulk case,
a relaxation of the consistency condition also occurs in
Brans-Dicke gravity. In the case studied here, however, the
absence of a negative tension brane for a consistent
compactification is due to the presence of the Brans-Dicke field,
not by some isolated bulk matter. In fact, in our analysis, all
extra bulk matter fields will be set to zero. In other words, we
show the existence of necessary conditions pointing to the fact that
the BD gravity can accommodate only branes with positive
tensions in a natural way. The presence of the BD scalar field
relax the consistency conditions. Besides, different from the scope
analyzed in Ref. \cite{INDY}, our demonstration is
completely model independent.

Going further in our analysis we allow the brane tension to be
time variable. Recently, the hypothesis that the Universe could be
better described by a variable brane tension has been raised
\cite{LAS}. In fact, keeping in mind the cosmological evolution of
the Universe, a variable brane tension seems to be an inevitable
condition. The application of the consistency conditions to simple
cases for variable brane tensions framework can be found in
Ref. \cite{PRI}. We shall implement the tools developed in
\cite{PRI} for the simplest case studied here. As we will see, a
dynamical brane tension may change its sign following the bulk cosmological
constant sign.

This paper is organized as follows: in the next Section we review
and extend the consistency conditions to the case of a Brans-Dicke
bulk. Then, in Section III, after particularizing to the
phenomenological interesting five-dimensional case, we show how the
scalar field acts in order to avoid ill defined scenarios. In
Section IV we study the case of time variable brane tensions and
in the last Section we summarize our results.

\section{Consistency conditions within Brans-Dicke gravity}

In this Section, we establish the basic mathematical tools which
will be used throughout this paper. In a sense, the consistency
conditions arise from an elegant  way of rewriting  the Einstein
tensor taking into account the presence of embedded branes in a
higher dimensional bulk. Their first derivations, employed to
specific cases, was developed in \cite{ANTES}. In this paper we
use the modern and general fashioned form developed in
\cite{CON,MAIS}. The full generalization of the consistency
conditions to the Brans-Dicke gravity case was obtained in
\cite{POS}. We shall outline, following the basic formulation of
reference \cite{MAIS}, the main steps fixing the notation used
here.

We start analyzing a D-dimensional bulk spacetime endowed with a
non-factorable geometry, whose metric is given by \be
ds^{2}=G_{AB}dX^{A}dX^{B}=W^{2}(r)g_{\alpha\beta}dx^{\alpha}dx^{\beta}+g_{ab}(r)dr^{a}dr^{b}\label{1},
\ee where $W^{2}(r)$ is the warp factor, assumed to be a smooth
integrable function, $X^{A}$ denotes the coordinates of the full
D-dimensional spacetime, $x^{\alpha}$ stands for the $(p+1)$
non-compact coordinates of the spacetime and $r^{a}$ labels the
$(D-p-1)$ directions in the internal compact space\footnote{As an
example, if $D=5$, $p=3$ and $W(r)=e^{-2k|r|}$ (being $k$ a
constant factor) we recover  the standard  Randall-Sundrum
model.}. Note that this type of metric encodes the possibility of
existing q-branes $(q>p)$ \cite{MAIS}. In this case, the $(q-p)$
extra dimensions are compactified on the brane and constitute part
of the internal space. This possibility is important in the hybrid
compactification models context, for instance \cite{JCAP}.

It is well known that the D-dimensional spacetime Ricci tensor can
be related with the brane Ricci tensor as well as with the
internal space partner by the equations \cite{CON} \be
R_{\mu\nu}=\bar{R}_{\mu\nu}-\frac{g_{\mu\nu}}{(p+1)W^{p-1}}\nabla^{2}W^{p+1},\label{2}
\ee and \be
R_{ab}=\tilde{R}_{ab}-\frac{p+1}{W}\nabla_{a}\nabla_{b}W,\label{3}\ee
where $\tilde{R}_{ab}$, $\nabla_{a}$ and $\nabla^{2}$ are
respectively the Ricci tensor, the covariant derivative and the
Laplacian operator constructed with the internal space metric
$g_{ab}$. $\bar{R}_{\mu\nu}$ is the Ricci tensor derived from
$g_{\mu\nu}$. Denoting the three curvature scalars by
$R=G^{AB}R_{AB}$, $\bar{R}=g^{\mu\nu}\bar{R}_{\mu\nu}$ and
$\tilde{R}=g^{ab}\tilde{R}_{ab}$, one sees that the traces of
equations (\ref{2}) and (\ref{3}) are given by \be
\frac{1}{p+1}\Big(W^{-2}\bar{R}-R^{\mu}_{\mu}\Big)=pW^{-2}\nabla
W\cdot \nabla W+W^{-1}\nabla^{2}W \label{4}\ee and \be
\frac{1}{p+1}\Big(\tilde{R}-R_{a}^{a}\Big)=W^{-1}\nabla^{2}W,\label{5}
\ee where $R^{\mu}_{\mu}\equiv W^{-2}g^{\mu\nu}R_{\mu\nu}$ and
$R^{a}_{a}\equiv g^{ab}R_{ab}$ (in such a way that
$R=R^{\mu}_{\mu}+R^{a}_{a}$). Now, being $\xi$ an arbitrary
constant, we have \be \nabla \cdot (W^{\xi}\nabla W)=W^{\xi+1}(\xi
W^{-2}\nabla W\cdot \nabla W+W^{-1}\nabla^{2} W)\label{6}.\ee
Finally, the combination of the Eqs. (\ref{4}), (\ref{5}) and
(\ref{6}) leads to \be \nabla \cdot (W^{\xi}\nabla
W)=\frac{W^{\xi+1}}{p(p+1)}[\xi\big(W^{-2}\bar{R}-R^{\mu}_{\mu}\big)+(p-\xi)\big(\tilde{R}-R^{a}_{a}\big)].\label{7}
\ee

This last equation encodes the basic statement of the consistency
conditions. The left-hand side (LHS) of Eq. (\ref{7}) vanishes upon
integration over the internal space while its right-hand side
(RHS) opens the possibility for the generalization to the
Brans-Dicke case since it expresses  the geometrical quantities in
terms of the scalar field. In fact, if the internal space is
compact without boundary or, equivalently, if the physical fields
are periodic at the extremities of the internal space interval,
the following relation must hold $\oint \nabla \cdot
(W^{\xi}\nabla W)=0$. Therefore, writing the RHS of Eq. (\ref{7}) in
terms of the spacetime stress tensor, one arrives at the
one-parameter $(\xi)$ family of consistency conditions. Each
choice of $\xi$ leads to an additional condition for the
consistency of the compactification scheme.

\subsection{The Brans-Dicke generalization of the consistency conditions}

In order to generalize to the Brans-Dicke gravity case we start by
remembering that the Einstein-Brans-Dicke bulk field equation is
given by
\begin{eqnarray} R_{MN}-\frac{1}{2}G_{MN}R&=&\left.
\frac{8\pi}{\Phi}T_{MN}+\frac{w}{\Phi^{2}}\Big(\nabla_{M}\phi\nabla_{N}\Phi-
\frac{1}{2}\nabla_{A}\Phi\nabla^{A}\Phi G_{MN}
\Big)\right.\nonumber\\&+&\left.\frac{1}{\Phi}\Big(\nabla_{M}\nabla_{N}\Phi-\frac{8\pi}{(D-1)+(D-2)w}TG_{MN}
\Big)\right.,\label{8}
\end{eqnarray}
where $T_{MN}$ is the matter stress-tensor, $T$ is the trace and
$w$ the BD parameter. We remark that the scalar part of the BD set
of equations, $\Box^{2}\Phi=\frac{8\pi}{(D-1)+(D-2)w}T$, was already taken
into account in the last term of the RHS of equation (\ref{8}). We
shall use the scalar equation again in the next Section after
doing some simplification in the calculations.
 From (\ref{8}), calling $T^{\mu}_{\mu}\equiv
W^{-2}g^{\mu\nu}T_{\mu\nu}$ $(T=T^{\mu}_{\mu}+T^{m}_{m})$, it is
possible to express $R^{\mu}_{\mu}$ and $R^{m}_{m}$ as
\begin{eqnarray}
R^{\mu}_{\mu}&=&\left.\frac{8\pi}{\Phi}\frac{1}{[(D-1)+(D-2)w]}\Big([D+w(D-p-3)-2]T^{\mu}_{\mu}-(1+w)(p+1)T^{m}_{m}\Big)
\right.\nonumber\\&+&\left.\frac{wW^{-2}}{\Phi^{2}}\nabla^{\nu}\Phi\nabla_{\nu}\Phi+\frac{W^{-2}}{\Phi}\nabla^{\nu}\nabla_{\nu}\Phi,\right.
\label{10}\end{eqnarray} and
\begin{eqnarray}
R^{m}_{m}&=&\left.\frac{8\pi}{\Phi}\frac{1}{[(D-1)+(D-2)w]}\Big([w(p-1)+p]T^{m}_{m}-(1+w)(D-p-1)T^{\mu}_{\mu}\Big)
\right.\nonumber\\&+&\left.\frac{w}{\Phi^{2}}\nabla^{m}\Phi\nabla_{m}\phi+\frac{1}{\Phi}\nabla^{m}\nabla_{m}\Phi.\right.
\label{11}\end{eqnarray} It is important to remark that when one
takes the limit $w\rightarrow \infty$, i. e. $\Phi\rightarrow
1/G_{N}$, the expressions $R^{\mu}_{\mu}$ and $R^{m}_{m}$ recover
the case analyzed in the General Relativity theory, as expected.

Replacing the Eqs. (\ref{10}) and (\ref{11}) in
(\ref{7}) one finds \begin{eqnarray} \nabla \cdot (W^{\xi}\nabla
W)&=&\left. \frac{W^{\xi+1}}{p(p+1)}\Bigg[\xi
W^{-2}\bar{R}+(p-\xi)\tilde{R}+\frac{8\pi}{\Phi}\frac{1}{[(D-1)+(D-2)w]}\right.\nonumber\\&\times&
\left.\Bigg(T^{\mu}_{\mu}[(p-\xi)(w+1)(D-p-1)-\xi\{D+w(D-p-3)-2\}]\right.\nonumber\\
&+&\left.T^{m}_{m}[\xi (w+1)(p+1)-(p-\xi)\{w(p-1)+p\}]\Bigg)\right.\nonumber\\
&-&\left.\frac{w}{\Phi^{2}}[\xi
W^{-2}\nabla^{\mu}\Phi\nabla_{\mu}\Phi+(p-\xi)\nabla^{m}\Phi\nabla_{m}\Phi]-\frac{1}{\Phi}[\xi
W^{-2}\nabla^{\mu}\nabla_{\mu}\Phi\right.\nonumber\\
&+&\left.(p-\xi)\nabla^{m}\nabla_{m}\Phi]\Bigg].\label{12}\right.
\end{eqnarray} Now, we shall write  explicitly the stress tensor partial
traces terms $T^{\mu}_{\mu}$ and $T^{m}_{m}$. To do that, we write
the bulk general stress tensor in the form \cite{MAIS} \be
T_{MN}=-\Lambda
G_{MN}-\sum_{i}T_{q}^{(i)}P[G_{MN}]_{q}^{(i)}\Delta^{(D-q-1)}(r-r_{i})+\tau_{MN},\label{15}
\ee where $\Lambda$ is the bulk cosmological constant,
$T_{q}^{(i)}$ is the $i^{th}$ q-brane tension with units given by
[energy/(length)$^{q}$], $\Delta^{(D-q-1)}(r-r_{i})$ is the
covariant combination of delta functions which positions the
brane\footnote{For a complete discussion about the expression of
$\Delta^{(D-q-1)}(r-r_{i})$ we refer the reader to Appendix of
the paper \cite{MAIS}.}, $P[G_{MN}]_{q}^{(i)}$ is the pull-back of
the bulk metric and any other matter contribution is due to
$\tau_{MN}$. From the Eq. (\ref{15}) it is simple to get \be
T^{\mu}_{\mu}=-(p+1)\Lambda+\tau^{\mu}_{\mu}-\sum_{i}T_{q}^{(i)}
\Delta^{(D-q-1)}(r-r_{i})(p+1),\label{16} \ee and \be
T^{m}_{m}=-(D-p-1)\Lambda+\tau^{m}_{m}-\sum_{i}T_{q}^{(i)}
\Delta^{(D-q-1)}(r-r_{i})(q-p)\label{17}.\ee

Inserting Eqs. (\ref{16}) and (\ref{17}) in
(\ref{12}) and using the identity $\oint \nabla \cdot
(W^{\xi}\nabla W)=0$ one obtains, after some algebra, the
following result
\begin{eqnarray}&&
\left. \oint W^{\xi+1}\Bigg[\xi
W^{-2}\bar{R}+(p-\xi)\tilde{R}-\frac{8\pi}{\Phi}\frac{1}{[(D-1)+w(D-2)]}\Bigg(-\sum_{i}T_{q}^{(i)}\Delta^{(D-q-1)}(r-r_{i})
\right.\nonumber\\&\times&\left.\Big[A(p+1)+B(q-p)\Big]
+A\tau^{\mu}_{\mu}+Bt^{m}_{m}-\Lambda\Big[A(p+1)+B(D-p-1)\Big]\Bigg)\right.\nonumber\\&-&\left.\xi
W^{-2}\Big[\frac{1}{\Phi}\nabla^{\mu}\nabla_{\mu}\Phi+\frac{w}{\Phi^{2}}\nabla^{\mu}\Phi
\nabla_{\mu}\Phi\Big]-(p-\xi)\Big[\mu\rightarrow
m\Big]\Bigg]=0\right. ,\label{18}
\end{eqnarray} where \be A\equiv (p-\xi)(w+1)(D-p-1)-\xi[D+w(D-p-3)-2],\nonumber
\ee \be B\equiv \xi(w+1)(p+1)-(p-\xi)[w(p-1)+p]\nonumber\ee are
constant parameters and the last term in (\ref{18}) denotes the
same parenthesis of the penultimate term, but with summation over
the internal space index $m$. The result encoded in Eq.
(\ref{18}) is exhaustive, in the sense that for every fixed $\xi$
it gives a constraint that must be obeyed for a consistent
compactification scheme in Brans-Dicke theory at a given
dimension. The equation (\ref{18}) itself, however, is not quite
useful since it is too general and does not give a specific
information about any particular model. Nevertheless, its
generality is exactly in what its importance resides: it
encapsulates in itself all such possible models.

In the next section we shall explore some possibilities which are
contained in the Eq. (\ref{18}), in particular, a
phenomenological interesting five-dimensional case. As we will
see, after some simplifications, it is possible to arrive at
strong and important constraints for the two branes scenario.

\section{Five-dimensional $AdS$ bulk case}

We shall apply in this Section the main result of the previous
section -- namely, Eq. (\ref{18}) -- to a particular case.
The scenario we would like to explore has the following set up:
two 3-branes embedded into a five-dimensional bulk spacetime
endowed with non-factorable geometry in Brans-Dicke gravity. This
type of scenario may be understood as the scalar-tensor analogous
to the Randall-Sundrum (RS) case. In the original RS model, one of
the basic, and necessary, characteristic of the brane tensions is
that it must be of opposite signs.

The main problem concerning  a negative tension object in the
compactification scheme is that such a brane is an inherently
unstable object\footnote{It is important to remark that some
specific frameworks where the negative tension brane is placed at
an orientfold plane may circumvent some objections concerning the
brane instability \cite{TRO}.}. In fact the bulk-brane system must
be a stable configuration solution of the gravitational field
equations, therefore without a negative tension object. The
problem with such type of branes may be understood as follows:
being the brane a (codimension one) submanifold embedded into a
higher dimensional manifold, it is always possible to project the
gravitational field equation from the bulk to the brane by, for
instance, the well known Gauss-Codazzi formalism \cite{WALD}. The
effective gravitational equation brings some specific signatures
from the extra dimensions with subtle but important departures
from the usual (four-dimensional) case. The effective Newtonian
constant on the brane inherits the tension brane sign. This
behavior is observed in both General Relativity and Brans-Dicke
gravities. Therefore, a negative tension brane induces a wrong
sign in the brane projected Newtonian constant.

From now on, let us assume a particular case where $D=5$ and
$p=q=3$ in order to mimic a RS scenario in the framework of the
Brans-Dicke gravity. As we will see, the existence of the scalar
field points to the possibility of the construction of a bulk-brane
structure without the necessity of any negative brane tension. With the chosen
dimensionality the constants $A$ and $B$ are simply given by
\begin{eqnarray} A(D=5,p=q=3)=3(1+w)-4\xi,\hspace{.5cm}
B(D=5,p=q=3)=6w(\xi-1)+7\xi-9.\label{20}
\end{eqnarray} Since this case has codimension one, the scalar of
curvature of the internal space (one line in the present case) is
zero, i. e., $\tilde{R}=0$. Then, discarding any additional matter
contribution for simplicity
$(\tau^{\mu}_{\mu}=\tau^{m}_{m}=0)$\footnote{Note the difference
between this approach and the one presented in Ref. \cite{INDY}.
In our case we do not use any additional bulk matter, working,
instead, within the BD gravity. It is possible to recover the case
studied in \cite{INDY} by setting $w\rightarrow \infty$ and
$\tau^{m}_{m}\neq 0$. As this must be clear from our approach, the
presence of additional bulk matter in General Relativity is also
responsible for a relaxing in the consistency conditions.}, Eq. (\ref{18}) reads
\begin{eqnarray}&& \left. \oint W^{\xi+1}\Bigg(\xi
W^{-2}\bar{R}+\frac{8\pi}{\Phi}\frac{1}{(4+3w)}\Bigg(-4A\sum_{i=1}^{2}T_{3}^{(i)}\delta(r-r_{i})-\Lambda(4A+B)
\Bigg)\right.\nonumber\\&-&\left.\xi
W^{-2}\Big[\frac{1}{\Phi}\nabla^{\mu}\nabla_{\mu}\Phi+\frac{w}{\Phi^{2}}\nabla^{\mu}\Phi
\nabla_{\mu}\Phi\Big]-(3-\xi)\Big[\mu\rightarrow
m\Big]\Bigg)=0,\right.\label{21}
\end{eqnarray} where $A$ and $B$ are given by Eq. (\ref{20}).
Now, two further simplifications. The first one comes from
experience. In trying to reproduce our universe one can set
$\bar{R}=0$ with an accuracy of $10^{-120}M_{Pl}$, where $M_{Pl}$
is the four-dimensional Planck mass \cite{CON}. Secondly, we shall
look at the consistency conditions focusing on the case where the
Brans-Dicke scalar field does not have dynamics on the brane, i.
e., $\nabla_{\mu}\Phi=0$. It may sounds as an oversimplification.
However, it is the most direct way not to contradict the
experimental gravitational bounds in the universe performed by a
brane \cite{EXP}. In other words, the simplest model consisting of
a Brans-Dicke gravity only in the bulk recovers  the (slightly
modified) General Relativity on the brane, in such a way that it is
compatible with the observational data.

Before reexpressing Eq. (\ref{21}) in terms of the above
simplifications it is worthwhile to mention that when we take the
General Relativity limit $w\rightarrow \infty$ $(\Phi=cte)$, the
$\Lambda$ coefficient $(4A+B)/(4+3w)$ reduces to $-2(\xi+1)$
which factorizes out the cosmological constant from the
consistency conditions for the $\xi=-1$ case, in close analogy to
the usual RS model presented in Ref. \cite{CON}.

Taking into account the above mentioned simplifications in Eq. (\ref{21}) we have \begin{eqnarray}&& \left. \oint
\frac{W^{\xi+1}}{\Phi}\Bigg(\frac{8\pi}{(4+3w)}\bigg(-4[3(1+w)-4\xi]\sum_{i=1}^{2}T_{3}^{(i)}\delta(r-r_{i})
-3[2W(\xi+1)+1-3\xi]\Lambda\bigg)
\right.\nonumber\\&-&\left.(3-\xi)\Big[\nabla^{m}\nabla_{m}\Phi+\frac{w}{\Phi}\nabla^{m}\Phi
\nabla_{m}\Phi\Big]\Bigg)=0.\right.\label{22}
\end{eqnarray} This last equation may be expressed in a more
suitable way with the aid of the scalar field equation of motion.
If the scalar field depends only on the extra dimensional
coordinate, then $\Box^{2}\Phi=\nabla^{m}\nabla_{m}\Phi$.
Furthermore, the energy-momentum trace is given by
$T=T^{\mu}_{\mu}+T^{m}_{m}=-4\sum_{i=1}^{2}T_{3}^{(i)}\delta(r-r_{i})-5\Lambda$,
in view of equations (\ref{16}) and (\ref{17}). Therefore, the
scalar equation becomes \be
\nabla^{m}\nabla_{m}\Phi=-\frac{8\pi}{4+3w}\Big(4\sum_{i=1}^{2}T_{3}^{(i)}\delta(r-r_{i})-5\Lambda
\Big).\label{23} \ee Inserting the Eq. (\ref{23}) into
(\ref{22}) we obtain, after a bit of algebra, the following
expression \be \oint \frac{W^{\xi+1}}{\Phi}\Bigg[\gamma
\sum_{i=1}^{2}T_{3}^{(i)}\delta(r-r_{i})+\eta \Lambda
-\frac{w(3-\xi)}{\Phi}\partial^{m}\Phi\partial_{m}\Phi\Bigg]=0,\label{24}
\ee where $\gamma$ and $\eta$ are constants given by
\begin{eqnarray} \gamma \equiv \frac{96\pi(\xi-w)}{(4+3w)}
, \hspace{.6cm} \eta \equiv
\frac{16\pi[6+2\xi-3w(\xi+1)]}{(4+3w)}.\label{25}
\end{eqnarray}

Now we are in a position to study the physical outputs of the
consistency conditions, by inserting some key values of the
parameter $\xi$. The most interesting choices are $\xi=3$, since
it simplifies the consistency condition eliminating the scalar
field derivative contributions, and $\xi=-1$ which eliminates the
overall warp factor term.

\subsection{The $\xi=3$ case}

Suppose $\xi=3$. In this case, the equations (\ref{24}) and
(\ref{25}) together give \be (3-w)\sum_{i=1}^{2}T_{3}^{(i)}\oint
\frac{W^{4}}{\Phi}\delta(r-r_{i})+2(1-w)\Lambda\oint\frac{W^{4}}{\Phi}=0.\label{26}
\ee Denoting by $W(r=r_{i})=W_{i}$ and $\Phi(r=r_{i})=\Phi_{i}$
the values of the warp factor and the scalar field on the i$^{th}$
brane, we have \be
(3-w)\sum_{i=1}^{2}\frac{W_{i}^{4}}{\Phi_{i}}T_{3}^{(i)}+2(1-w)\Lambda\oint\frac{W^{4}}{\Phi}=0.\label{27}
\ee Implementing the $AdS$ bulk case $\Lambda<0$ in Eq.
(\ref{27}), it simply reads \be
\frac{(3-w)}{2(1-w)}\sum_{i=1}^{2}\frac{W_{i}^{4}}{\Phi_{i}}T_{3}^{(i)}=|\Lambda|\oint\frac{W^{4}}{\Phi}.\label{28}
\ee The remarkable characteristic of the above expression is that
it shows a necessary condition pointing to the viability of a consistent compactification scheme
with only {\it positive} tension branes, avoiding then ill defined
scenarios.

The typical functional form of warp factors solving the hierarchy
problem are $W\sim \exp(-f(r))$, being $f(r)$ determined by the
gravitational field equations. However, as shown in Eq.
(\ref{28}), the important term is given by a ``coupling'' \,
between the warp factor and the scalar field. In fact, the
proposition of braneworld models in the framework of scalar-tensor
gravitational theories resides in the fact that  the scalar field
composes the warp factor. In this sense, it is not surprising that
a term as $W^{4}/\Phi$ is determinant to the establishment of the
compactification scheme. In other words, the term $W^{4}/\Phi$ can
be seen as a composite warp factor, which may act as a tool to
solve the hierarchy problem {\it allowing}, at the same time,
the presence of only positive tension branes. As an aside remark
we stress that for flat spacetimes ($W=1$, everywhere) the
presence of negative tension branes may be also unnecessary.

\subsection{The $\xi=-1$ case}

Let us now suppose $\xi=-1$. With such specification the $\gamma$
and $\eta$ constants read \be
\gamma=-\frac{96\pi(1+w)}{4+3w},\label{29}\ee \be
\eta=\frac{64\pi}{4+3w}\label{30}, \ee and Eq. (\ref{24}) gives \be
\frac{24\pi(1+w)}{4+3w}\sum_{i=1}^{2}\frac{T_{3}^{(i)}}{\Phi_{i}}+
\frac{16\pi}{4+3w}|\Lambda|\oint \frac{1}{\Phi}+w\oint
g^{rr}[\partial_{r}(ln\Phi)]^{2}=0.\label{31}\ee Rewriting the
above equation in a more suitable way, we have \be \oint
g^{rr}[\partial_{r}(ln\Phi)]^{2}=-\frac{8\pi}{4+3w}\Bigg\{\frac{2|\Lambda|}{w}\oint\frac{1}{\Phi}+
\frac{3(1+w)}{w}\sum_{i=1}^{2}\frac{T_{3}^{(i)}}{\Phi_{i}}\Bigg\}.\label{32}
\ee The left-hand side of the equation (\ref{32}) is positive,
therefore the positive tension branes are possible, being $\oint 1/\Phi <0$ and
\ba \Bigg|\oint\frac{1}{\Phi}\Bigg|>\frac{3(1+w)}{2|\Lambda|}\sum_{i=1}^{2}\frac{T_{3}^{(i)}}{\Phi_{i}}\label{vamola},\ea
which can be satisfied by a typical power-law behavior of the scalar field, i. e., $\Phi \sim r^{-a}$.

\section{A time variable tension case}

In this Section we shall study an interesting characteristic of a
time variable tension toy model. In Ref. \cite{LAS}, a
complete solution for a time tension variable braneworld highly
compatible with the observable cosmological symmetries is shown.
Recently, the time variation of brane tensions was analyzed with
the aid of the consistency conditions \cite{PRI}. In this Section,
we shall point out an interesting characteristic arising from a
simple toy model based on the time variable tension branes: in
an AdS bulk, it is possible for the brane to change the sign of
its tension at some point of the time evolution.

In Ref. \cite{PRI}, it is shown that the time variation of the brane
tension may be accomplished by the following replacement \be
T_{3}^{(i)}\longrightarrow \tilde{T}_{3}^{(i)}\equiv
T_{3}^{(i)}+\kappa^{(i)}\partial_{t}T_{3}^{(i)},\label{33}\ee in
Eq. (\ref{24}). In (\ref{33}) $\kappa^{(i)}$ is a
positive constant with units of (energy)$^{-1}$, keeping
$\tilde{T}_{3}^{(i)}$ with units of (energy)/(length)$^{3}$. This
simple prescription ignores the possible high derivative
corrections in the time variation and does not concern  the
mechanism under which the tension becomes variable. However, this
toy model does not preclude the existence of non-linear
effects in the brane tension. We would like to remark here that it
might be possible to achieve a more realistic variation to the brane
tension by looking at some cosmological upper bounds of the local
rate of change of the gravitational `constant' \cite{NEG}, since
both --- the brane tension and the gravitational `constant'--- are
related by dimensional reduction of the bulk gravitational
equations. Here, we shall keep our analysis to the following toy
model, stressing a curious behavior of the brane tension dynamics
in an AdS bulk.

The development of the consistency conditions formulae with the
replacement given by (\ref{33}) can be done in the very same way
of what we have done in the previous section. From Eq.
(\ref{24}) it is easy to see that \be
\sum_{i=1}^{2}\frac{\tilde{T}_{3}^{(i)}W_{i}^{\xi+1}}{\Phi_{i}}=\frac{w(\xi-3)}{\gamma}\oint
W^{\xi+1}g^{rr}[\partial_{r}(ln\Phi)]^{2}-\frac{\eta}{\gamma}\Lambda\oint\frac{W^{\xi+1}}{\Phi}.\label{ane}
\ee For simplicity, we may assume that the tension of the hidden
brane behaves independently of the visible one, i. e., that its
tension is simply given by $T_{3}^{hid}\sim
\exp(-t/\kappa^{hid})$, such that $\tilde{T}_{3}^{hid}=0$. The
characteristics which we are going to arrive at do not depend strongly on
this assumption. We shall keep it in order to emphasize our point
in a clear way. Keeping it in our minds, Eq. (\ref{ane}) results in
the following differential equation \be
\dot{T}_{3}^{vis}+\frac{1}{\kappa^{vis}}T_{3}^{vis}+\frac{1}{\kappa^{vis}}F(r)=0,\label{34}
\ee where $\dot{T}_{3}^{vis}=\frac{dT_{3}^{vis}}{dt}$ and $F(r)$
is just the right-hand side of Eq. (\ref{ane}) multiplied by
the $\Phi_{vis}/W^{\xi+1}_{vis}$ factor. The solution of
(\ref{34}) is given by \be
T_{3}^{vis}(t)=Ce^{-t/\kappa^{vis}}-\frac{1}{\kappa^{vis}}\frac{\Phi_{vis}}{W^{\xi+1}_{vis}}\Bigg[\frac{w(\xi-3)}{\gamma}\oint
W^{\xi+1}g^{rr}[\partial_{r}(ln\Phi)]^{2}-\frac{\eta}{\gamma}\Lambda\oint\frac{W^{\xi+1}}{\Phi}\Bigg],\label{35}
\ee where $C$ is a positive integration constant (with units of
tension). We remark that taking the limit $w\rightarrow \infty$
($\Phi$ constant) one arrives again at the simplest case studied
in General Relativity \cite{PRI}, as expected. If
one is willing to accept the current experimental bound \cite{EXP} of the
Brans-Dicke parameter for the bulk, the first term is dominant
over the second one, where the cosmological constant is present.
Nevertheless it may not be the case, i. e., the Brans-Dicke theory
may have another ``status'' in the bulk. Apart of this, the
choices $\xi\geqslant3$ show that $T_{3}^{vis}$ must change its sign
at some time, provided that $\phi_{vis}$ and $W_{vis}$ have the
same sign on the brane. Particularly, for $\xi=3$ the relation
between the brane tension and the cosmological constant sign
appears explicitly. For this case, the Eq. (\ref{35}) reduces
to \be
T_{3}^{vis}(t)=Ce^{-t/\kappa^{vis}}-\frac{2|\Lambda|}{\kappa^{vis}}\frac{(1-w)}{(3-w)}\frac{\Phi_{vis}}{W^{4}_{vis}}
\oint\frac{W^{4}}{\Phi},\label{36}\ee and the tension changes its
sign in the specifical time \be
\bar{t}=\kappa^{vis}\ln\Bigg\{\frac{\kappa^{vis}C}{2|\Lambda|}\frac{(3-w)}{(1-w)}\frac{W^{4}_{vis}}{\Phi_{vis}}\Bigg(\oint
\frac{W^{4}}{\Phi}\Bigg)^{-1}\Bigg\}.\label{37}\ee Obviously the
value $\bar{t}$ is lower for bigger absolute values of the bulk
cosmological constant.

We shall interpret the general picture encoded in Eq.
(\ref{35}) as follows: although the contribution of the first term
of the right-hand side may be positive or negative, the sign of
the second one is always negative for an $AdS_{5}$ bulk slice.
Therefore (apart of the relative values of the two terms), it acts
dynamically pulling the brane's tension to a negative value. It
may be useful to make a rough analogy between the time variation
of the brane tension and a classical mechanical system, thinking
that $T_{3}^{vis}(t)$ is the position of a particle according to
the time variation. The equation (\ref{34}) has a ``friction''
term given by $\dot{T}_{3}^{vis}$  which is responsible for a
damping in the $T_{3}^{vis}$ curve. It is not difficult to see
that the conservative terms of the equation of motion (\ref{34})
can be derived by the following ``Lagrangian'' independent of the
velocities \be
L(T_{3}^{vis},t)=-\frac{F(r)}{\kappa}T_{3}^{vis}-\frac{1}{2\kappa}(T_{3}^{vis})^{2},\label{38}
\ee in such a way that \be \frac{d}{dt}\Bigg(\frac{\partial
L}{\partial \dot{T}_{3}^{vis}}\Bigg)-\frac{\partial L}{\partial
T_{3}^{vis}}=-\dot{T}_{3}^{vis},\label{39}\ee and the equation
(\ref{34}) can be recovered. The energy associated to the
Lagrangian is given by \be E=\dot{T}_{3}^{vis}\frac{\partial
L}{\partial
\dot{T}_{3}^{vis}}-L=\frac{T_{3}^{vis}}{\kappa}\Bigg(\frac{T_{3}^{vis}}{2}+F(r)\Bigg).\label{40}
\ee Deriving Eq. (\ref{40}) with respect to the time and
using (\ref{34}) one gets \be
\frac{dE}{dt}=-(\dot{T}_{3}^{vis})^{2}, \label{41}\ee so there is
energy loss due to the friction term, as expected. Let us analyze
a little further the equation (\ref{41}) exploring the possibility
of a minus sign for the $F(r)$ term, \be
E=\frac{T_{3}^{vis}}{\kappa}\Bigg(\frac{T_{3}^{vis}}{2}-|F(r)|\Bigg).\label{42}
\ee Note that if $T_{3}^{vis}/2>|F(r)|$ the energy remains
positive. On the other hand, being $0<T_{3}^{vis}/2<|F(r)|$ we
have a negative energy $E<0$. However, if $T_{3}^{vis}<0$ the
energy of the mechanical system is always positive. Therefore, a
negative $T_{3}^{vis}$ (for $F(r)<0$) is widely consistent with
the positivity of the energy; in fact it guarantees a positive
energy system. As an aside remark, note that if $\Lambda=0$ in Eq.
(\ref{36}) then the energy of the system never becomes
negative (see (\ref{42})), i. e., the time $\bar{t}$ above which
the tension changes its sign tends to infinity, as can be seen
from (\ref{37}).

This analogy, although useful, should be made with great care. The
``Lagrangian'' (\ref{39}) does not define a consistent dynamical
system in the strict Hamiltonian sense. It is easy to see that the
constraints of the system defined by (\ref{39}) are such that the
``Euler-Lagrange'' equation (\ref{34}) seems to be inconsistent
\cite{DIRAC}. Note, moreover, that by taking this analogy quite
seriously one faces the difficult interpretation of a damped
system (due to the friction term) whose Lagrangian does not depend
on the velocities (therefore with vanishing acceleration). It is
possible that with a more complete replacement in the relation
(\ref{33}), taking into account second order derivatives of
$T_{3}^{vis}$, we can set a consistent Hamiltonian dynamical
system. We shall investigate it in a future work. With this
warning in mind, the rough analogy expressed above is useful to
get some physical insight about the shift of the brane tension
sign.

\section{Concluding Remarks}

In this paper we review and extend the consistency conditions
analysis for branes embedded into a bulk respecting Brans-Dicke
gravity. It is shown that this scenario allows (by furnishing necessary conditions)
the existence of only positive tension branes in a five-dimensional AdS bulk.
This seems to be very attractive, since negative tension branes
are inherently unstable objects. This possibility, in practice,
arises due to the relaxation of the consistency conditions,
coming from the presence of the Brans-Dicke field, i. e., the studied
consistency conditions are softened by the addition of the scalar field.
Qualitatively, the presence of the scalar field engenders a
coupling as $W^{r}/\phi$ (being $r$ a real number) on the brane,
which can be interpreted as a new (composite one) warp factor,
``cancelling out'' the negative contribution of the AdS bulk
cosmological constant (see equation (\ref{27}), for instance).

Going further, we studied a time variable brane tension toy model.
It was shown that a dynamical tension brane can change its sign,
aligning  with the sign of the bulk cosmological constant. This is
an example of how interesting the dynamics of the brane tension
can be, i. e., even in a simplified toy model, as the present one.
Even though simple, it is possible to find out an unusual and rich
behavior for the tension time variability.

We would like to remark that the results presented, mainly in
Section III, point out to the necessity of including the scalar
field in the five dimensional bulk in order to obtain a completely
compatible compactification scheme. We can say that this is our
main result in this paper since all is based on the fact
that gravity is of scalar-tensor nature. Some future research
lines may follow from such statement. As stressed in Section III,
the Brans-Dicke scalar field depends only on the extra (out of
brane) dimension. Perhaps, a more complete scenario may arise by
allowing the scalar field to vary with the brane coordinates. In
this case, the use of the consistency conditions should be
complemented by a more complete set of inputs, approaching a
specific model, since the presence of terms accounting for the
brane coordinates variation of the scalar field complicates the
analysis (see equation (\ref{21}), for instance). Apart of that,
one must be careful in order to not contradict the experimental
bounds to the Brans-Dicke parameter.

To summarize, the results presented in this paper suggest that a
more complete scenario --- contemplating also time variable
tension branes --- deserves special attention, since it may be the
source of unusual and unexplored properties of braneworld
scenarios.

\section*{Acknowledgments}

It is a pleasure to thank profs. M. C. Bertin and Roldão da Rocha
for quite useful discussions. J. M. Hoff da Silva thanks to
FAPESP-Brazil (PDJ 2009/01246-8) for financial support and
Universidade Federal do ABC (UFABC), where this work was partially
done. M. C. B. Abdalla and M. E. X. Guimarães thank CNPq for a
support. We also thanks to the anonymous JHEP referee for useful comments
about this work.

\end{document}